# Situational Awareness based Risk-Adapatable Access Control in Enterprise Networks

Brian Lee, Roman Vanickis, Franklin Rogelio and Paul Jacob
*Software Research Institute, Athlone Institute of Technology, Athlone, Ireland*
{blee,pjacob}@ait.ie, ,{frogelio,rvanickis}@research.ait.ie

Keywords: Risk based Access Control, RAdAC, Zero-Trust Networking, Security Situational Awareness,

Abstract: As the computing landscape evolves towards distributed architectures such as Internet of Things (IoT), enterprises are moving away from traditional perimeter based security models toward so called "zero trust networking" (ZTN) models that treat both the intranet and Internet as equally untrustworthy. Such security models incorporate risk arising from dynamic and situational factors, such as device location and security risk level risk, into the access control decision. Researchers have developed a number of risk models such as RAdAC (Risk Adaptable Access Control) to handle dynamic contexts and these have been applied to medical and other scenarios. In this position paper we describe our ongoing work to apply RAdAC to ZTN. We develop a policy management framework, FURZE, to facilitate fuzzy risk evaluation that also defines how to adapt to dynamically changing contexts. We also consider how enterprise security situational awareness (SSA) - which describes the potential impact to an organisations mission based on the current threats and the relative importance of the information asset under threat - can be incorporated into a RAdAC scheme.

## 1 INTRODUCTION

The enterprise computing landscape has evolved considerably over the last decade with the emergence of technologies such as cloud computing and mobile devices, including the "bring your own device", (BYOD) phenomenon. These trends are likely to be amplified even further in coming years with the increasing deployment of Internet of Things (IoT) applications. Also, over the same period, there has been a dramatic escalation in both the number and sophistication of security-attacks on business (Broadhurst et. al. 2014).

This combination puts pressure on the traditional perimeter based security model toward the application of so called "zero trust networking" (ZTN) models that treat the intranet with the same degree, i.e. lack, of trust as the internet (Forrester Research, 2013). Access control decisions are now made based on the degree of trust in the user and the device, irrespective of whether the user is inside or outside the internal network (Ward and Beyer, 2014). Furthermore the level of access assigned to a device or user may change over time and the access control system is able to dynamically infer the current trust level by consulting various data sources and make decisions accordingly.

One approach to the capability of an access control system to adapt dynamically to changing context is Risk-Adaptable Access Control (RAdAC), (McGraw, 2009). The key concept of RAdAC is the trade-off, or balance, between *operational need* and *security risk* when making access control decisions - in some conditions operational need will override security risk and access to otherwise restricted resources will be granted. RAdAC was originally proposed for highly variable environments such as those in the military but is increasingly seen applied in other situations characterised by dynamic contexts such as medical emergencies and the IoT.

Similarly, dynamic access control for enterprise networks has been considered for some time, (Fernandez, 2006) and is becoming ever more appropriate as the variety and sophistication of security attacks and the complexity of these networks increases (Farroha and Farroha, 2012). The types of factors used for access control in enterprise ZTN scenarios includes user and device as well as situational factors such as location and access history, (Ward and Beyer, 2014; Vensmer and Kiesel, 2012).

In most applications of RAdAC described to date the focus has been *relaxation* of the access controls in order to maximise the benefit for the organisation by granting temporary access to restricted objects. In certain cases however the goal might be to *restrict* access as e.g. in the case of heightened security situation (Fernandez, 2006) and it is this particular issue we address here.

In particular we consider how enterprise *security situational awareness (SSA)* - which describes the potential impact to an organisations mission based on the current threats and the relative importance of the information asset under threat - can be incorporated into a RAdAC scheme. We extend the RAdAC model of Kandala, Sandhu and Bhamidipati (2011), in this direction and define a framework to manage risk based access control in zero trust networks.

The remainder of the paper is structured as follows: Section 2 describes the background concepts. Section 3 describes our proposed RAdAC approach to zero-trust networking. Section 4 describes related work while conclusions and future work are outlined in Section 5.

## 2 BACKGROUND

### 2.1 Access Control

For many years the dominant model of access control has been Role Based Access Control (RBAC). RBAC is centred around the notion of authorisation via groups or roles i.e. access rights are allocated to a group rather than an individual.

However RBAC has limitations for the evolving enterprise computing landscape. In particular it does not very well support multi-factor decisions such as those based on location, special training received etc. as well as rank in the organisation, (Hu et al. 2013) Nor are it's primarily static role definitions well suited for dynamic access control decisions. As a result a more dynamic access control approach has been proposed i.e. Attribute Based Access Control (ABAC), (OASIS 2013). ABAC decisions are based on the use of subject and object attributes and enables very flexible access control capabilities. ABAC also introduces the concepts of obligations and access based on environment conditions. Obligations are additional actions required by the subject as part of pre- or post-conditions of granting authorisation. Environment conditions enable the inclusion of context specific criteria into the access control decision-making such as the location of the user, the state of emergency in a hospital theatre etc. ABAC systems are beginning to see significant deployment in the enterprise including use for dynamic access control based on environment conditions, (Osborne et al. 2016; Vensmer and Kiesel, 2012).

While ABAC significantly increases the dynamic capability of access control systems it does not include the capability to factor risk into authorisation decision-making. To this end McGraw (2009) proposed an extension of ABAC to manage dynamic risk assessment - the Risk Adaptable Access Control (RAdAC) concept. According to McGraw (2009) – "RAdAC incorporates a real time, probabilistic determination of security risk into the access control decision rather than just using a hard comparison of the attributes of the subject and object as in traditional models". RAdAC is based on assessing the trade-off between *security risk* and *operational need*. The definition of these terms is very much context and organisation dependent though possible inputs to security risk determination could include e.g. trust in user or device, the current threat level, sensitivity, or criticality to the mission, of the requested

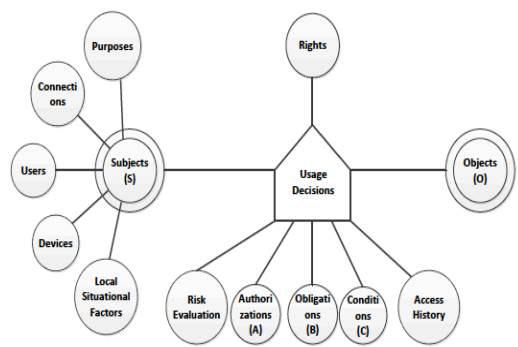

Figure 1: UCON Model with RAdAC Components, (from Kandala, Sandhu and Bhamidipati 2011).

information, role of the user etc. (Farroha and Farroha, 2012).

Kandala, Sandhu and Bhamidipati (2011) define an abstract model for RAdAC in terms of the "Usage Connection" (UCON) access control approach– shown in Figure 1. UCON is an extended access control approach that seeks to unify both traditional access control i.e. access at the *start* of the transaction with the need for on-going control of access to the object *during* the transaction – what UCON terms *decision continuity*. This latter property is a significant addition to RAdAC as it allows adaptation to changing environment conditions. In Figure 1 above the subject concept has been decomposed into a number of components i.e. *users*, *devices*, *connections* and *purposes*. The usage/access control decision process is shown to include a *Risk Evaluation* component as well predicate/rule-based components for *Authorisation*, (based on the attributes) *Obligations* (as for ABAC) and *Conditions* (captures the environmental conditions*)*. Usage access decision-making is based on all three rule/predicate components i.e. authorisation, obligation and conditions and all can be evaluated pre, during or post the session.

## 2.2 Zero Trust Networking

There is an increasing move away from the traditional perimeter-based security approach both in the enterprise and in cloud computing (Ward and Beyer 2014). This trend has been dubbed "Zero Trust Networking" (Forrester, 2013). The key element of this approach is to treat the internal network as untrusted to the same degree as the Internet. This usually entails partitioning the internal network into a number of network segments or *zones* each of which houses different functions and information. Each zone may have a different *trust level* that indicates the importance of the assets housed within the zone (Gontarczyk et al. 2015). In order to access an asset, a subject's trust level assignment must be equal to or greater than the zone's minimum trust level (Osborne et al. 2016). Traffic between zones is thus restricted by firewalls in accordance with the overall access control policy.

Access control in ZTN is transaction based and dynamic i.e. a decision is made for each access request and firewall rules are updated as a result of each decision. Access control decisions are based on user and device attributes as well as environment attributes such as location. Deployments include enterprise security e.g. Google's BeyondCorp (Osborne et al. 2016) as well as campus (Vensmer and Kiesel 2012) and cloud computing security (Giannakou et al. 2016).

## 2.3 Mission Impact Modelling

As outlined above environmental factors that are critical parameters in determining situational risk include those relating to the current threat level and the business or mission criticality of the related assets. This approach can also play a part in the access control system to determine if access should be granted based on the criticality of the resource being accessed and the current threat level.

At the heart of this approach is the "*mission dependency graph*" (Holspopple and Yang 2013; Innerhofer–Oberperfler and Breu 2006; Jakobson 2011; Watters et al. 2009). The mission dependency graph maps the organisations strategic mission objectives to the IT assets by defining, and prioritising, the dependencies between both via the different layers of the graph i.e. business objectives are realised (depend on) by business functions and processes which in turn are, completely or partly, realised (depend on) by IT capabilities and assets. The edges in the graph record a causal relationship/influence between two nodes and may be assigned a weighting to reflect the extent of that influence as in Holspopple and Yang (2013) and Watters et al. (2009). These values are typically assigned through consultation with key business and IT personnel responsible for the various processes and components that constitute the graph.

A mission graph is used for two complementary purposes. Firstly it is used to assign to each IT asset an *asset criticality* dependent on the particular missions or business goals it supports. This is done by trickling the mission objective priority *down* the graph to the individual asset. This feature is used to help analysts improve situational awareness as discussed above. Secondly the degree of severity of a particular threat to a particular asset can be evaluated to determine the risk to the mission objectives by percolating the risk *up* the graph. This risk assessment feature can be used for off-line what-if analyses and it also forms the basis for our situational awareness risk based access control.

Another consideration that must be taken into account is how the base impact -the impact to the information asset at the root of the mission tree that is the subject of a threat - is calculated. Watters (2009) uses vulnerability severity as a base for his calculation as does Jakobson (2011). Holspopple and Yang (2013) simply assume a measure of impact without specifying how it may be calculated while Innerhofer–Oberperfler and Breu (2006) do not address the issue at all.

## 3 FURZE

Whilst Kandala, Sandhu and Bhamidipati (2011) have provided a sound abstract model for "UCON-ised" RAdAC a number of significant research questions remain open around the practical deployment of RAdAC. Our area of interest is to explore enforcement architecture issues for the application of RAdAC in ZTN through the development of a policy enforcement framework that we designate FURZE (Fuzzy Risk Framework for ZTN).

Specific issues we wish to investigate include i) the definition of a policy management architecture to include the on-going monitoring needed to enable decision continuity in ZTN access, ii) the design of an access control policy language to specify and manage such updates, iii) the design a risk evaluation function based on fuzzy logic to enable probabilistic access control decisions to be taken and iv) to investigate how the proposed approach works for enterprise security situational awareness.

The starting point for FURZE is strongly hinted by Kandala, Sandhu and Bhamidipati (2011) in Figure 1. This contains a number of elements that form the functional, logical and linguistic basis for the proposed policy framework for ZTN shown in Figure 2. The architecture is broadly based on the XACML policy framework (OASIS 2013). Access requests are received via a Policy Enforcement Point (PEP) as e.g. a WiFi base station. The Context Handler coordinates the access control process including decision continuity handling. The Environment Evaluation contains plug-in components that convert session and other relevant factors into attributes that can be used for risk evaluation e.g. through the use of rules. Examples include SSA (discussed below), operational need and locations - for a simple access control case these components could simply default to session attributes. The Risk Evaluation Function and Access Decision Function jointly act as a Policy Decision Point (PDP) to control access while the subject and object attributes are stored in a management database. The Firewall Provisioning module converts generic firewall rules to device specific filtering rules for appropriate firewalls.

In FURZE the application of decision continuity imposes the need on the access control function to maintain session state information so that access control can adapt to reflect situational or other influencing factors that change the balance between operational need and security risk and trigger policy re-evaluation. Risk assessment is made as part of the initial authorisation predicate evaluation and, possibly, subsequently as part of either an authorisation or condition predicate evaluation that in turn has been triggered by some event during the session.

The definition of the FURZE *policy language* will be based on i) subject and object entities and their attributes and ii) policy rules formed from authorisation and condition predicates. The subject entity is decomposed into a number of components that including e.g. *device* (from which the user requests access), *purpose,* which reflects operational need and is the underlying reason for the users access request, *local situational factors* impacting the individual user e.g. location etc. Objects define the services, data and other resources users wish to access. Global situational factors such as enterprise security risk are evaluated as condition predicates independent of any individual user.

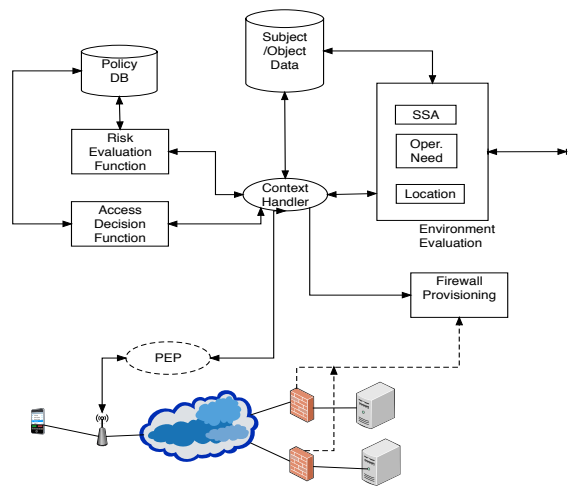

Figure 2: FURZE Framework.

For the ZTN specific case the BeyondCorp framework gives broad hints at policy factors while the DynFire framework also suggests some specific factors to consider as access control *credentials* (Vensmer and Kiesel 2012). Similarly previous work on policy management for firewall access control (e.g. Zhang et al. 2007) can contribute to the development of the policy language. The principal challenge here for FURZE is to combine selected elements of previous work with language constructs to incorporate i) decision continuity and ii) risk evaluation. Decision continuity will be handled in the Context Handler.

Risk evaluation is at the heart of RAdAC access control languages, as RAdAC must include the capability to reckon with uncertainty. The proposed approach for risk evaluation in FURZE – through the Risk Evaluation component- will be based on the use of fuzzy logic inferencing. Fuzzy logic is a mathematically sound approach to draw unambiguous conclusions from vague or imprecise information and subjective if-then rules (Ni, Bertini and Lobo, 2010) and has been widely used for decision-making under uncertainty including risk assessment. Furthermore as Ni shows, fuzzy logic is flexible can be used to model a variety of access control approaches. In FURZE we propose to leverage existing fuzzy logic language developments – specifically the use of the Fuzzy Control Language (FCL) (IEC 61131-7 2000). FCL is a domain specific language (DSL) created for use in fuzzy control application in industrial automation. It enables the definition of domain relevant fuzzy concepts including linguistic variables, term sets and rule sets. The principal challenge is to wrap or embed the FCL risk evaluation into the overall FURZE policy language. This includes not just seamless language syntax but also satisfactorily resolving DSL implementation issues.

Our proposed approach to evaluate mission impact (via the SSA component) is based on the use of Fuzzy Cognitive Maps (FCM) (Kosko, 1986) to model both the information asset threat impact and the mission graph. FCM is a graph based approach to modelling systems in which the concepts that capture different aspects of the behaviour of the system as represented as nodes while weighted edges capture interactions between concepts that represent the dynamics of the system. We base our threat modelling on the approach proposed by Jakobson (2011) but extend it to deal with different aspects of impact relating to the security trio of confidentiality, integrity and availability. We will use Rule Based Fuzzy Cognitive Maps (RB-FCM) to model the mission tree as it allows us to represent a richer set relationship than standard FCM (Carvalho and Tome, 1999).

## 4 RELATED WORK

In relation to RAdAC the present work can be regarded as an extension of the earlier conceptual work (Kandala, Sandhu and Bhamidipati (2011; Farroha and Farroha 2012) toward implementation of an enforcement architecture.

Google has adopted a ZTN approach for access control that, as described, seems partly similar to the work defined here (Ward and Beyer 2013; Osborne et al. 2016). However they have not described the policy language, risk management or decision continuity implementation in detail. Vensmer and Kiesel (2012) describe Dynfire, an access control policy management framework for ZTN applied to a university campus that encompasses a number of the ideas described in our work. However it does not include either risk management nor decision continuity.

Giannoku et al. (2016) describe AL-SAFE a ZTN access control implementation for cloud computing –however they also do not include policy language, risk management or decision continuity aspects.

Firewall policy management is a mature research area and a number of authors have described have described efforts in this direction (Zhang et al. 2007; Lobo, Marchi and Provetti 2012; Cisco 2010). However none of these authors have described access control risk management as part of their contributions.

A number of researchers have modelled the impact of information security attacks on the organisation business and mission (Jakobsson 2011; Innerhoffer—Oberperfler and Breu 2006; Holspopple and Yang, 2013). However none have included mission impact assessment as part of the access control decision continuity process.

Cheng et al. (2007) developed a fuzzy logic approach to access control risk assessment that proposed the gradation of security risk as services of levels between "allow" and "deny" where each level has an associated risk mitigation countermeasure. Ni, Bertini, and Lobo (2010) investigated the applicability of fuzzy inference for risk based access control and concluded that the approach was flexible and scalable. Our work most closely matches that of Ni, Bertini, and Lobo (2010) as we also propose the use of fuzzy inference for risk calculation – however our scope of research is somewhat wider.

## 5 CONCLUSION

We have described a novel approach to risk based access control for ZTN networking. The key element of our contribution is a policy control framework based on UCON that facilitates situational awareness based decision continuity and which embeds a fuzzy inference component to make the actual risk estimates. Further we have shown how asset criticality based on mission importance can be used as part of the situational awareness process. Future areas of research are to implement and evaluate the proposed approach for a ZTN scenario and to investigate its applicability to virtualised networking in cloud computing and software-defined networking (SDN).

## ACKNOWLEDGEMENTS

This paper has received funding from the European Union's Horizon 2020 research and innovation programme under grant agreement 700071